\documentclass[aps,pre,showpacs]{revtex4}
\usepackage{graphicx}
\usepackage{epsfig}
\begin{document}
        
\title{Exact Solution for a 1-dimensional model for Reptation}

\author{Andrzej Drzewi\'nski$^{(a)}$ and J.M.J. van Leeuwen$^{(b)}$}
\affiliation{$^{a}$Czestochowa University of Technology, 
Institute of Mathematics and Computer Science,
ul.Dabrowskiego 73, 42-200 Czestochowa, Poland}
\affiliation{$^{b}$Instituut-Lorentz, University of Leiden, P.O.Box 9506, \\
2300 RA Leiden, the Netherlands}

\date{\today} 
  
\begin{abstract}
We discuss the exact solution for the properties of the recently introduced
``necklace'' model for reptation. The solution gives the drift velocity,
diffusion constant and renewal time for asymptotically long chains. Its properties
are also related to a special case of the Rubinstein-Duke model in one dimension.
\end{abstract}


\maketitle

\section{Introduction}

The most common lattice model for reptating polymers is the Rubinstein-Duke model
\cite{Rubinstein,Duke}. The major drawback of this model is, that it has sofar defied an
exact solution. Therefore it is natural to look for models that are in the same spirit,
but which are more open to analytic rather than numerical methods. Recently Guidoni et al.
\cite{Terranova} introduced and analyzed a 1-dimensional model for reptation,
to which we shall refer as the ``necklace model''.
It is a chain that moves through the exchange of beads and vacancies
along a line. The purpose of this note is to show that the model permits an exact solution
by means of the Matrix Product Expansion, due to Derrida et al. \cite{Derrida2}. Also the
spectrum of the Master Operator can be derived, leading an exact expression for the renewal
time. In addition we relate the properties with those of a special 1-dimensional variant
of the Rubinstein-Duke model with hernia creation and annihilation.

\section{The Necklace Model}

The necklace model is a string of $N+1$ beads located on a line of points.
The beads are  either neighbors or nearest neighbors.
In the latter case there is an unoccupied lattice
site (a vacancy) between the beads. The beads are not allowed to occupy the same lattice
point and two consecutive vacancies are forbidden in order to ensure the integrity
of the chain. The possibility of vacancies is an expression of the elasticity of the chain.
The internal beads hop with by exchanging with a vacancy. The two end-beads can also
exchange with a vacancy from outside. Clearly the number of beads is conserved, but not the
number of vacancies, which enter and leave at the ends and migrate through the chain.
The stationary state of the system follows from the Master Equation, governing this
stochastic motion.

The key to the exact solution is to focus on the motion of the vacancies rather than on that of
the beads. There are $N$ positions for the vacancies available, since each vacancy must be
surrounded by beads. Each of the $N$ positions can be occupied by a vacancy or not.
The state of the chain of beads is fully determined by the occupation distribution of the
vacancies. So we consider the chain as an open system for vacancies, which hop with the
rates derived from the motion of the original beads. Their hopping is constrained by the rule
that two vacancies can not occupy the same position. It would mean that the two beads
surrounding such a double vacancy would be separated by two vacancies.

Transfering the motion for the beads to the vacancies we arrive at the following rules
\begin{itemize}
\item An internal vacancy can hop to a neighboring empty position with rate $p_c$
to the right and $p'_c$ to the left.
\item The exchange of a vacancy with outside goes with rates $p_a$ for entering
and $p'_a$ for leaving at the left hand end of the chain.
\item On the right hand end, the rates are $p_b$ for leaving and $p'_b$ for entering.
\end{itemize}
Note that we have primed rates for the hops to the left and unprimed
for the motion to the right. The asymmetry between left and right can be attributed to a
driving field on the beads, inducing a bias $B$ for a hop to the right and a bias $B^{-1}$
for a leftward hop. These biases may derive from a charge on the beads, which is influenced
by an electric field. Then we have for the ratios the relations
\begin{equation} \label{1}
{p'_a \over p_a} = {p'_b \over p_b} = {p'_c \over p_c} = B^2.
\end{equation}
For the solubility of the model it is not neccessary to assume these ratios and only when we
discuss the properties in more detail, we will use this physical restriction.
The above given rates are more general than those used by Guidoni et al., in particular we
allow for a finite driving field.  For our exact solution, it is however
important that the hopping rules for the internal vacancies are uniform along the chain.

\section{The Master Equation}

The state of the chain can be represented by $N$ variables $\tau_i$, which assume the values
1 and 0, where 1 viz. 0 corresponds to the presence viz. absense of a vacancy.
The probability distribution $P(\tau_1, \cdots, \tau_N)$ of the stationary state follows from
the Master Equation
\begin{equation} \label{a0}
{\cal M} \, P(\tau_1, \cdots, \tau_N) = 0,
\end{equation}
with $\cal M$ the Master Operator. It contains the usual gain term consisting of all
transitions which increase the probability $P(\tau_1, \cdots, \tau_N)$ and a loss term
which contains all transitions out of the configurations $(\tau_1, \cdots, \tau_N)$. The
Master Operator is the sum of $N+1$ operators representing the action of the beads
\begin{equation} \label{a1}
{\cal M}= \sum^N_{i=0} {\cal M}_i.
\end{equation}
As first example consider the bead at the left hand of the chain, leading to ${\cal M}_0$.
It acts on the $\tau_1$ dependence of $P (\tau_1, \cdots, \tau_N)$, as it only influences
the existence of a vacancy on the first position. It is given by the expression
\begin{equation} \label{a2}
{\cal M}_0 P (\tau_1 \cdots ) = (1-\tau_1) [p'_a P (1 \cdots) - p_a P(0 \cdots)] +
\tau_1 [p_a P (0 \cdots) - p'_a P(1 \cdots)].
\end{equation}
The first 2 terms refer to the case where a vacancy is absent from the first possible position.
It has a gain and a loss term. Similarly  the last 2 terms refer to the case where a vacancy
is present. The expression can be shortened to
\begin{equation} \label{a3}
{\cal M}_0 P (\tau_1 \cdots ) = (1 - 2 \tau_1) [p'_a  P (1 \cdots) - p_a P(0 \cdots)].
\end{equation}
In the same way the other parts of the Master Operator are expressed as
\begin{equation} \label{a4}
\left\{ \begin{array}{rcl}
{\cal M}_j P (\cdots \tau_j, \tau_{j+1} \cdots) & = & (\tau_j -\tau_{j+1})
[p'_c P (\cdots 0,1 \cdots) - p_c P (\cdots 1,0 \cdots)]\\*[2mm]
{\cal M}_N P (\cdots \tau_N) & = & (1 - 2\tau_N) [p_b  P (\cdots 1) - p'_b P(\cdots 0)].
\end{array} \right.
\end{equation}
This form of the Master Operator shows that it has an eigenvalue 0, as summation over all
$\tau_i$ yields zero for all individual terms, implying conservation of probability.
The left eigenvector is  constant for all configurations, but the right eigenvector is the
non-trivial probability distribution of the stationary state.

\section{Matrix Product Expansion for the Stationary State}

The idea of the matrix product expansion is to represent the probability distribution
$P (\tau_1, \cdots, \tau_N)$  as product of matrices
\begin{equation} \label{b1}
P (\tau_1, \cdots, \tau_N) = Z^{-1}_N \langle W | \prod^N_{j=1}
[\tau_j {\cal D} + (1- \tau_j) {\cal E} ]| V \rangle.
\end{equation}
As we shall see in a minute, ${\cal D}$ and  ${\cal E}$ are sort of creation and annihilation
operators, represented by infinite dimensional matrices and $|V \rangle$ and
$ \langle W|$ are states in this space. The normalization is given by
\begin{equation} \label{b2}
Z_N = \langle W | ({\cal D} + {\cal E})^N | V \rangle.
\end{equation}
The matrices ${\cal D}$ and ${\cal E}$, as well as the states $\langle W|$, and
$| V \rangle$ are to be determined such that (\ref{b1}) is the stationary state of the
Master Equation. By insertion of (\ref{b1}) we find the action of the ${\cal M}_j$
on the factors
\begin{equation} \label{b3}
\left\{ \begin{array}{rcl}
{\cal M}_0 [\tau_1 {\cal D} + (1-\tau_1) {\cal E}] & = &
(1 -2 \tau_1)(p'_a {\cal D} - p_a {\cal E}),  \\*[2mm]
{\cal M}_j [\tau_j {\cal D} + (1 - \tau_j) {\cal E}]
[\tau_{j+1} {\cal D} + (1-\tau_{j+1}) {\cal E}] & = &
(\tau_j - \tau_{j+1})(p'_c{\cal E} {\cal D} - p_c {\cal D} {\cal E}), \\*[2mm]
{\cal M}_N [\tau_N {\cal D} + (1-\tau_N ) {\cal E}] & = &
(1 -2 \tau_N)(p_b  {\cal D} - p'_b  {\cal E}).
\end{array} \right.
\end{equation}
Then we use the identity
\begin{equation} \label{b4}
(\tau_j - \tau_{j+1}) ({\cal D} + {\cal E}) =
(1 - 2 \tau _{j+1}) [ \tau_j {\cal D} + (1 - \tau_j) {\cal E}]
-(1 - 2 \tau _j) [ \tau_{j+1} {\cal D} + (1 - \tau_{j+1}) {\cal E}]
\end{equation}
and impose the condition
\begin{equation} \label{b5}
p'_c {\cal E} {\cal D} - p_c {\cal D} {\cal E} = \zeta [{\cal D} + {\cal E}],
\end{equation}
in order to rewrite the expression for the internal reptons as
\begin{equation} \label{b6}
\begin{array}{c}
{\cal M}_j [\tau_j {\cal D} + (1 - \tau_j) {\cal E}][\tau_{j+1} {\cal D} +
(1-\tau_{j+1}) {\cal E}] = \\*[2mm] \zeta (1 - 2 \tau _{j+1}) \,[ \tau_j {\cal D} + (1 - \tau_j)
{\cal E}]   - \zeta (1 - 2 \tau_j)\,[ \tau_{j+1} {\cal D} + (1 - \tau_{j+1}) {\cal E}].
\end{array}
\end{equation}

The ``commutation'' relation (\ref{b5}) defines the properties of the ${\cal D}$ and
${\cal E}$ matrices. The states $|V \rangle$ and $\langle W |$ are fixed by the relations
\begin{equation} \label{b7}
\langle W |(p'_a {\cal D} - p_a {\cal E})  = \langle W | \zeta,\quad \quad
(p'_b {\cal E} - p_b {\cal D}) |V\rangle =\zeta |V \rangle
\end{equation}
Then it is possible to recombine in (\ref{b3}) the factors again to probabilities
\begin{equation} \label{b8}
\left\{ \begin{array}{rcl}
{\cal M}_0 P (\tau_1, \cdots, \tau_N) & = & \zeta (1 - 2 \tau_1) P (\tau_2 ,\cdots,\tau_N)
\\*[2mm]
{\cal M}_j P (\tau_1, \cdots, \tau_N) & = & - \zeta (1 - 2 \tau_j) P (\tau_1, \cdots, \tau_{j-1},
\tau_{j+1}, \cdots, \tau_N) \\*[2mm]
 & + & \zeta  (1 - 2 \tau_{j+1} ) P (\tau_1 \cdots, \tau_j, \tau_{j+2}, \cdots, \tau_N) \\*[2mm]
{\cal M}_N P (\tau_1, \cdots, \tau_N) & = & -\zeta  (1 - 2 \tau_N) P(\tau_1, \cdots, \tau_{N-1})
\end{array} \right.
\end{equation}
The parameter $\zeta $ is arbitrary; it has been introduced to facilitate the normalization.
The sum over all these relations vanishes, which shows that (\ref{b1}) is the
stationary state of the Master Equation.

Relation (\ref{b5}) can be clarified by the substitution
\begin{equation} \label{b9}
{\cal D}  =  \displaystyle 1 +b^\dagger, \quad \quad \quad
{\cal E} =\displaystyle  1+  b
\end{equation}
This yields for the creation and annihilation operators
$b^\dagger$ and $b$ the $q$-deformed commutation relation
\begin{equation} \label{b10}
b b^\dagger - q b^\dagger b = 1 - q,
\end{equation}
provided that we fix $\zeta$ and $q$ as
\begin{equation} \label{b11}
\zeta = p'_c - p_c, \quad \quad \quad q = p_c/p'_c.
\end{equation}
The advantage of the new operators is that their spectrum has been thoroughly investigated.
They have similar properties  as the usual creation and annihilation operators for the
harmonic oscillator. Note that for $q=1$ the operators seem to commute, but it is easy to
rescale $b^\dagger$ and $b$ such that the right hand side of (\ref{b10}) becomes 1, as often
is done \cite{Blythe}.

The power of the representation is that it gives the probability distribution for
all chain lengths $N$. The properties of the matrices ${\cal D}$ and ${\cal E}$ and
the states  $\langle W|$ and $| V \rangle$ are independent of $N$. It requires, however,
quite a bit of formal manipulations to retrieve the chain properties from the general
expression (\ref{b1}).

The relations (\ref{b5}) and (\ref{b7}) were mentioned by Derrida et al. \cite{Derrida2}.
They were related to the $q$-deformed algebra by Sasamoto \cite{Sasamoto} and Blythe et al.
\cite{Blythe}, leading to explicit formulae. Unfortunately, the case that we see as most
physical (relations (\ref{a1}) and all intrinsic mobilites the same), does not fall in the wide
range of parameters already treated. However, the technique that they employ, can be used
to deal with our case. As it is not our aim to marginally extend the exact solutions
to an even wider regime, we refer to \cite{Sasamoto,Blythe} for details.

\section{The Drift Velocity}

The phase diagram as function of the parameters is governed by the magnitudes of the
net hopping rates
\begin{equation} \label{c1}
\Delta p_a = p'_a -p_a, \quad \quad \Delta p_b = p'_b -p_b,
\quad \quad \Delta p_c = {1 \over 2} (p'_b -p_c)
\end{equation}
The smallest is the limiting factor. If the input on the left hand side is small, a dilute phase
results. If the output at the right hand side is small, a dense phase will form. If the
throughput in the bulk of the chain is the limiting factor, a maximum current phase appears,
with a constant density of vacancies in the bulk.

For illustration we discuss the case (\ref{a1}), with hopping rates only influenced by the driving
field with bias $B$. Then we are in the maximum current phase. We restrict ourselves to the
most interesting property: the drift velocity. The drift of bead $j$, in a specific configuration,
is given by
\begin{equation} \label{c2}
v_j = p'_c P(\cdots 0,1 \cdots) - p_c P(\cdots 1,0 \dots).
\end{equation}
The first term is the probability to jump to the right and the second for that to the left.
Remember that the beads jump in the opposite direction of the vacancies. Comparing this
with relation (\ref{a4}), we have the identity
\begin{equation} \label{c3}
\tau_j {\cal M}_j P(\tau_1, \cdots, \tau_N) = \tau_j (1 - \tau_{j+1}) \, v_j.
\end{equation}
By summing over the variables $\tau_j$ and $\tau_{j+1}$, the factor in front of $v_j$ gives 1
and one finds for the average drift
\begin{equation} \label{c4}
\langle v_j \rangle = \langle \tau_j {\cal M}_j \rangle.
\end{equation}
This average can be expressed, with the aid of (\ref{b8}) as
\begin{equation} \label{c5}
v = Z^{-1}_N \langle W | ({\cal D} + {\cal E})^{N-1} \ V \rangle = {Z_{N-1} \over Z_N} \zeta.
\end{equation}
We have dropped the index $j$ as the result is obviously independent of $j$. It has to,
since there is no accumulation of beads in the stationary state.  So (\ref{c5})
is the expression for the drift velocity, relating it to the normalization factors $Z_N$.
The proportionality of the drift to  $\zeta = p'_c - p_c$ is simply a matter of times scales:
when the rates go up, the drift follows proportionally.

The $Z_N$ are expressed in terms of what is called the position operator $b^\dagger+b$
of the $q$-deformed harmonic oscillator.
\begin{equation} \label{c6}
{\cal D} + {\cal E} = 2+ b + b^\dagger.
\end{equation}
The spectrum of $b^\dagger+b$ has been investigated in detail \cite{Sasamoto}. It can be
represented as
\begin{equation} \label{c9}
(b + b^\dagger)| \theta \rangle = 2 \cos \theta | \theta \rangle
\end{equation}
The spectrum is continous: $0 \leq \theta \leq \pi$. So we get
\begin{equation} \label{c10}
Z_N = \int^\pi_0 d \theta w(\theta) [2(1+ \cos \theta)]^N \, \langle W |\theta \rangle
\langle \theta | V \rangle
\end{equation}
Here $w (\theta)$ is a function that enters in the closure relation for the states
$| \theta \rangle$ (see \cite{Sasamoto}). For large $N$ the power of $1 + \cos \theta$
gets strongly peaked around $\theta =0$ and one can apply the saddle point method to
find the leading term. We put $\theta =0$ in all non-singular terms yielding
\begin{equation} \label{c11}
Z'_N \simeq z(q) 4^N N^{-3/2}
\end{equation}
where the factor $z(q)$ is independent of $N$ and contains amongst others the factor
$\langle W |0 \rangle \langle 0 | V \rangle$. Fortunately this factor drops out in the
drift velocity.
\begin{equation} \label{c12}
v \simeq {1 \over 4 } \, (p'_c - p_c)
\end{equation}
The drift vanishes of course when the bias $B \rightarrow 1$, but its leading term
does not decay with the power $N^{-1}$ as does the drift in the RD-model.

The diffusion coefficient is usually defined in the limit that the driving field vanishes
$B \rightarrow 1$. Using the Einstein relation one has to divide by a factor $N$ and this
leads to the (dimensionless) value
\begin{equation} \label{c13}
D \simeq {1 \over 4 N}
\end{equation}
Relations (\ref{c12}) and (\ref{c13}) show that we are in the maximum current regime, which
is not limited by the input and output at the ends of the chain.

\section{Small Chains}

The asymptotically leading behavior is of course the most interesting result, showing the
strength of the Matrix Product Expansion. However, it is also worth while to see that the whole
behavior is determined by the commutation relation (\ref{b10}). To illustrate this point
we confine ourselves to the physical restriction (\ref{1}). Then the equations for
$ \langle W |$ and $ | V \rangle $ reduce to
\begin{equation} \label{d1}
\langle W | \, b^\dagger = q \langle W | \, b, \quad \quad \quad
b \, | V \rangle = q b^\dagger | V \rangle.
\end{equation}
Note that
$\langle W | = \langle V |$ since the first relation (\ref{d1}) is the conjugate of the second.

As first example of playing with commutation relations consider
\begin{equation} \label{d2}
\langle V | (b + b^\dagger) | V \rangle = q \langle V | (b^\dagger + b ) | V \rangle
\end{equation}
where we used (\ref{d1}) to interchange $b$ and $b^\dagger$ by applying  $b$ to
$| V \rangle $ and $b^\dagger$ to $\langle V |$. As $q \neq 1$ the average has to
vanish. This holds also for all odd powers of $b + b^\dagger$. The second example
is the average of $b^\dagger b$, for which we first use the relations (\ref{d1})
\begin{equation} \label{d3}
\langle V | b b^\dagger | V \rangle = q^{-2} \langle V| b^\dagger b | V \rangle
\end{equation}
But the commutation relation (\ref{b10}) can also be used
to convert the right hand side of (\ref{d3}) to the left hand side expression. From
these two relations one deduces
\begin{equation} \label{d4}
 \langle V | b^\dagger b | V \rangle = {q^2 (1-q) \over (1-q^3)} \, \langle V | V \rangle
\end{equation}
All other averages of 2 creation or annihilation operators directly follow,
with (\ref{d1}), from this expression. So the current of the 2 link system equals
\begin{equation} \label{d5}
v_{N=2} = {2 (1+q +q^2) \over 5 + 6 q + 5 q^2} (B-B^{-1})
\end{equation}
an expression, which of course, can also be obtained by solving the probability
distribution from the Master Equation. By this technique it will be a long and hard road to
get to the behavior at large $N$.

\section{The Renewal Time}

Another interesting quantity is the renewal time. It is defined as the slowest time of decay
towards the stationary state and it follows from the spectrum of the Master Operator as the
eigenvalue with the smallest negative real part. Clearly the corresponding state decays the
slowest and the renewal time is the inverse of the gap in the spectrum. The matrix product
representation does not lead to the full spectrum of the Master Operator; it only gives the
stationary state eigenfunction. On the other hand the renewal is usually defined in the
fieldless case for which the Master operator becomes hermitian (or symmetric in our case).
The terms of the Master Operator can be expressed in terms of the operators $a^\dagger_j$,
creating a vacancy, and $a_j$ annihilating a vacancy (see \cite{Sartoni}).
\begin{equation} \label{e1}
\left\{\begin{array}{rcl}
{\cal M}_0 & = & B (a_1 - a^\dagger_1 a_1) + B^{-1} (a^\dagger_1 -a_1 a^\dagger_1),\\*[2mm]
{\cal M}_j & = &   B (a^\dagger_j a_{j+1} - a_j a^\dagger_j a^\dagger_{j+1} a_{j+1})
+ B^{-1} (a_j a^\dagger_{j+1} - a^\dagger_j a_j a_{j+1} a^\dagger_{j+1}),\\*[2mm]
{\cal M}_0 & = & B^{-1} (a_N - a^\dagger_N a_N) + B (a^\dagger_N -a_N a^\dagger_N).
\end{array} \right.
\end{equation}
The $a_j$ are hard core boson operators, equivalent with spin $1/2$ operators, via the relations
\begin{equation} \label{e2}
a^\dagger = (\sigma^x + i \sigma^y)/2, \quad \quad \quad a = (\sigma^x - i \sigma^y)/2.
\end{equation}
with the consequence
\begin{equation} \label{e3}
a^\dagger a = (1 + \sigma^z)/2, \quad \quad\quad a a^\dagger =(1 - \sigma^z)/2,
\end{equation}
where the $\sigma$'a are the Pauli Matrices. The Master operator, expressed in these spin
operators, is an unusual hamiltonian since it is non-hermitian. However, for the undriven
system $B=1$, it turns into the well known ferromagnetic Heisenberg chain.
\begin{equation} \label{e4}
{\cal M} = {1 \over 2} \sum^{N-1}_{j=1} [\, {\bf \sigma}_j {\bf\sigma}_{j+1} -1] + \sigma^x_1
+ \sigma^x_N -2.
\end{equation}
Due to the magnetic field in the $x$ direction on the boundaries of the chain, it is profitable to
work in a basis of eigenstates of $\sigma^x$. These are the symmetric (spin up) and the
anti-symmetric (spin down) combinations of a vacancy and its absence.
\begin{equation} \label{e5}
|\uparrow \, \rangle = 2^{-1/2}\, [\,|1 \rangle + |0 \rangle\,], \quad \quad \quad
|\downarrow \, \rangle = 2^{-1/2}\, [\,|1 \rangle - |0 \rangle\,].
\end{equation}
The ``groundstate'' (being the highest in the spectrum) of the chain is the state with all spins
directed in the $x$ direction, which has an eigenvalue 0. In vacancy language this is the
state in which all configurations have the same probability.

As the hamiltonian (\ref{e4}) conserves the number of spins in the $x$ direction
the spectrum breaks up into sectors with a given number of spins up. The smallest excitation
from the ``groundstate'' is in the sector with one spin down. Let $x_n$ be the value of the
state with the down spin at the position $n$. Then we have the set of equations
\begin{equation} \label{e6}
\left\{ \begin{array}{rcrcrcrcr}
-(3+ \lambda) x_1 & + &             x_2 &   &     &         &   &             = & 0 \\*[1mm]
	      x_1 & - & (2+ \lambda) x_2 & + & x_3 &         &   &             = & 0 \\*[1mm]
\cdots            &   &                 &   &     &         &   &             = & 0 \\*[1mm]
		  &   &                 &   & x_{N-1} & - & (3+ \lambda) x_N & = & 0 \\*[1mm]
\end{array} \right.
\end{equation}
The eigenvalue spectrum is readily evaluated and one finds
\begin{equation} \label{e8}
\lambda (k) = - 2 (1 - \cos (\pi k/N)), \quad \quad \quad k = 1, 2, \cdots, N.
\end{equation}
Thus the gap in the spectrum is given for $k=1$ with
\begin{equation} \label{e9}
\lambda (1) = - 2 (1 - \cos (\pi/N)).
\end{equation}
The other branches can be evaluated similarly, but lead to larger negative eigenvalues.

\section{The Rubinstein-Duke Model}

The Rubinstein-Duke (RD) model is in a way complementary to the necklace model. In
the RD model the dynamical elements are reptons, blobs of monomers of the
order of persistence length. A polymer chain is a string of reptons on a lattice with the
constraint that reptons are either in the same cell or in neighboring cells. There are two types
of links between reptons: the {\it slack} links connect two successive reptons in the same cell
and the {\it taut} links those in neighboring cells. A repton can hop to a
neigboring cell if it does not leave an empty cell behind. So the basic mechanism of motion
is the interchange of taut and slack links. The elasticity of the chain is due to
storing length through occupation of cells with more than one repton.
The integrity of the chain is guaranteed by the requiring that a connected tube of cells
is traced out by the chain. The natural embedding is in a lattice of dimension $d>1$, but
Duke \cite{Duke} showed that it suffices to study the projection on the driving field direction.
Thus the dimension $d$ becomes a parameter and the links are characterised by
a 3-valued variable: $y_i=0$ for a slack link, $y_i=1$ for a taut link
in the direction of the field and $y_i=-1$ for a taut link opposite to the field. The
parameter $d$ influences the ratio of slack to taut links, but not the universal properties
of the chain e.g. the exponents by which the renewal time and the drift velocity depend on
the chain length. So the model is often studied for $d=1$ although the 1-dimensional version
is rather artificial \cite{Widom}.

Sartoni and van Leeuwen \cite{Sartoni} noticed that the 1-dimensional version of the RD model
could be related to a model with two types of particles. Consider the superposition of two
non-interacting 1-dimensional systems of particles.
The particles are called $+$ viz. $-$ and they hop according to the same rules as the vacancies
in the necklace model model. So they are not allowed to occupy the same site, but,
since they are non-interacting, a $+$ particle may occupy the same site as a $-$ particle.
Therefore each site can be in 4 states: empty, occupied by a $+$ or a $-$ or double occupied,
which we indicate by a $\pm$. In total there are $4^N$ configurations. As the two systems
do not interact, the probabilities of the combined system are the product of the
probabilities of the two systems. Each of them can be treated with the Matrix Product
Expansion and yields identical expressions for e.g. the drift velocity.

Having the freedom to choose the transition rates of the two systems independently, we
drive the $+$ particles in one direction and the $-$ particles in the opposite
direction with equal strength. Then the $+$ particles are identified with the links $y_i =1$
of the RD model and the $-$ particles with the links $y_i=-1$. The $0$ and $\pm$ state
are both mapped on a slack link $y_i=0$.
This gives a contraction of the $4^N$ configurations of the superposition to the $3^N$
configurations of the RD model. Knowing the probability distribution of the combined system
one can construct the transition rates in the Master Equation for the contracted system.

It  will in general be a quite involved calculation. We illustrate it by a simplifying
assumption, which is motivated by the following consideration. By reversing the driving
field, the role of particles and vacancies are interchanged. The system is invariant under
a transformation that maps the particles of one component onto the vacancies of the other
component. In this map an empty site, which is a vacancy state for both components, maps
onto a doubly occupied site. Thus, as we will see, it is reasonable to equate
the probabilities for an empty and a doubly occupied site.

\begin{figure}[h]
\centering
\includegraphics[width=10cm]{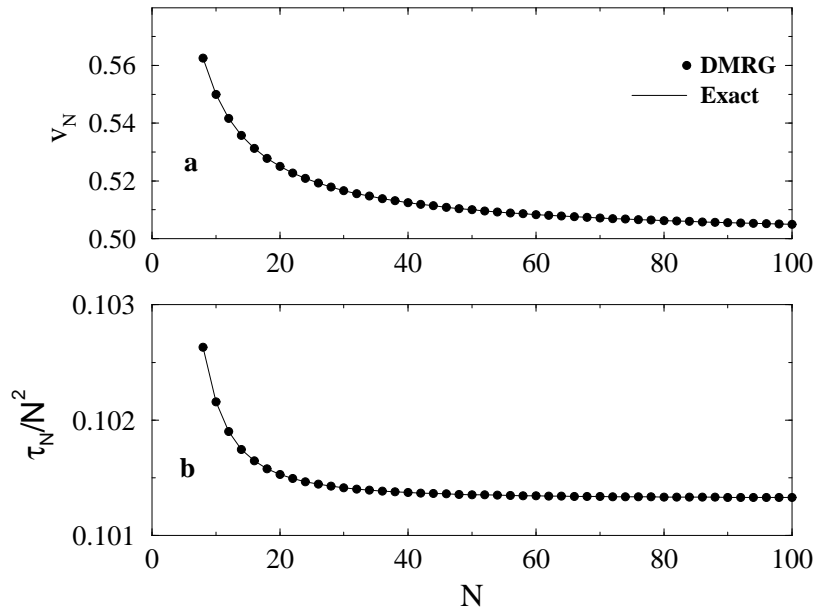}
\caption{The drift velocity for the RD model with hernias as a function of
the length of the chain, together with the expression (\ref{f1}) for the necklace
model (a). Part (b) gives the renewal time of the RD chain with the expression
(\ref{e9}) deduced from the necklace model.}
\label{compar}
\end{figure}

Before we list the result for the transition rates of this constructed RD model, we note that
the underlying combined system has a feature not present in the usual RD models. If we
have a succession of two taut links, it can be a combination of the $0$ and $\pm$ state.
In that case, one of the particles of the $\pm$ state can move to the empty site, thereby
creating a pair of opposite taut links. This is called the creation of a hernia. It is
not incorporated in the usual RD model, since it is a modification of the tube
(consisting of the set of taut links). Similarly the opposite process: a $+$ and $-$ particle
meeting each other on the same site, is the annihilation of a hernia. With this in mind we
come to the following list of transitions.
\begin{itemize}
\item Transitions of a slack-taut combination. The slack link corresponds to the 2 states
of the $+$ and $-$ particle system. Each of them can interchange the taut and slack link,
so the transition rate is the same.
\item Transitions of a slack-slack combination. This corresponds to 4 states of the
particles. Two of them cannot move: the $0 , 0$ and the $\pm , \pm$
combination. The two others can create a hernia. So the hernia creation rate is 1/2.
\item Hernia annihilation. A hernia corresponds to a unique particle state and it may
develop into a slack-slack pair in two ways: the $+$ particle may move or the $-$
particle. So the hernia annihilation rate is 2.
\item A slack end-link. It has 2 particle configurations, $0$ and $\pm$,
and both may move to the taut position,
either by creating a particle in the state $0$ or by annihilation of a particle
in the state $\pm$. So the transition rate is 1.
\item A taut end-link. It is a unique particle state and it can transform itself in two
ways in a slack state: by annihilation of the particle or by creation a particle of the
other kind. So it has a transition rate 2.
\end{itemize}
Thus the particle system maps onto a chain with asymmetric transition rates for the
end reptons and for the creation and annihilation of hernias. The other moves are the same
as in the RD model.
The model with the above listed transition rules will demonstrate the same behavior as
the superposition of the $+$ and $-$ particles system with opposite driving fields.
Although the map is based on a global and not a local symmetry, it is correct in the
weak driving field limit. We have calculated independently the diffusion coefficient
and the gap of this model by means of the DMRG method\cite{DMRG}.
Fig. \ref{compar}a compares the drift velocity with the expression (\ref{c13}),
refined with finite size effects (as can be deduced from \cite{Sartoni}).
\begin{equation} \label{f1}
v_N = {N+1 \over 2N}.
\end{equation}
Note that the drift velocity of the RD model is twice the value of the necklace model, since
there are two systems  of particles moving independently.

Fig. \ref{compar}b does the same for the renewal time and compares it with
the expression (\ref{e9}).
As one observes the agreement is perfect. Looking into the numbers one has to conclude that
the differences can be made arbitrary small by making the DMRG calculation more accurate.
It proves also that the DMRG method is amazingly accurate in calculating reptating polymer
chains.

\section{Discussion}

We have given an exact solution for the stationary state of the necklace model introduced by
Guidoni et al. \cite{Terranova}. Also an exact expression is given for the low lying
excitations (gap). This model mimics the reptation of a chain in a 1-dimensional
system. It is not difficult to formulate the model in higher dimensions, but a chain in
higher dimensions cannot be reconstructed from the position of the vacancies.
Moreover, if such a system is driven by a field, the transition rates not only depend
on the vacancy distribution, but also on the direction of the connecting links.
So one arrives at a model of the same complexity as the RD model.
The model shows in $d=1$ not the characteristics of the slow reptation behavior. The
drift velocity for long chains approaches a constant rather than decaying with the inverse
powers of the length. The reason is that the model has no obstacles which slow down the
drift and the diffusion.

The solution can also be used for a special RD model in $d=1$, with the possibility of creation
and annihilation of hernias. This shows the importance of the hernias in $d=1$. Whereas it
is believed that hernia creation and annihilation is of minor importance in higher dimension,
it plays a decisive role in $d=1$. Without the hernias as a move, the $+$ taut links and the
$-$ taut links block each other and they are driven towards each other. A hernia annihilation
followed by a creation of a pair of taut links in the opposite order, allows them to pass each
other. So the hernias, which are very abundant in $d=1$, effectively remove the obstacles,
which are characteristic for the RD model.

\bigskip

{\bf Acknowledgments} One of the authors (JMJvanL) is indepted to Martin Depkin for
introducing him into the literature on the $q$-deformed algebra as a means to analyze
the Matrix Product Expansion.

\end{document}